\documentclass[10pt]{article}
\usepackage[top=2cm,left=2cm,right=2cm]{geometry}
\usepackage[numbers,square]{natbib}

\usepackage[english]{babel}
\usepackage[latin1]{inputenc}
\usepackage{hyperref}
\usepackage{amssymb}
\usepackage{amsfonts}
\usepackage{amsmath}
\usepackage{fancyhdr}
\usepackage{epsfig}
\usepackage{graphics}
\usepackage{subfigure}
\usepackage{rotating}
\usepackage{lineno}
\usepackage{pstricks}
\usepackage{color}

\let\a = \alpha    \let\b =\beta

\newcommand{\muu}{\mu_{1}}
\newcommand{\nuu}{\nu_{1}}
\newcommand{\ku}{k_{1}}
\newcommand{\xu}{x_{1}}

\newcommand{\mud}{\mu_{2}}
\newcommand{\nud}{\nu_{2}}
\newcommand{\kd}{k_{2}}
\newcommand{\xd}{x_{2}}

\newcommand{\mut}{\mu_{3}}
\newcommand{\nut}{\nu_{3}}
\newcommand{\kt}{k_{3}}
\newcommand{\xt}{x_{3}}

\newcommand{\muq}{\mu_{4}}
\newcommand{\nuq}{\nu_{4}}
\newcommand{\kq}{k_{4}}
\newcommand{\xq}{x_{4}}

\newcommand{\beq}{\begin{equation}}
\newcommand{\eeq}{\end{equation}}
\newcommand{\beqa}{\begin{eqnarray}}
\newcommand{\eeqa}{\end{eqnarray}}
\newcommand{\bea}{\begin{eqnarray}}
\newcommand{\eea}{\end{eqnarray}}

\newcommand{\nn}{\nonumber}

\newcommand{\pd}{\partial}

\begin{document}
\begin{center}
\vspace{4.cm}
{\bf \large Higher Order Dilaton Interactions in the Nearly Conformal Limit \\ of the Standard Model}

\vspace{1cm}
{\bf Claudio Corian\`{o}, Luigi Delle Rose, Carlo Marzo and Mirko Serino}

\vspace{1cm}

{ Dipartimento di Matematica e Fisica \\ 
Universit\'a del Salento \\ and \\ INFN Lecce, Via Arnesano 73100 Lecce, Italy\\}
\vspace{0.5cm}

\begin{abstract}
We investigate the Standard Model in the nearly conformal limit, in which conformal symmetry is broken only by the dilatation anomaly, through a hierarchy of anomalous Ward identities for the divergence of its dilatation current.  In this approximation, the identities allow to extract the coupling of the dilaton to the trace anomaly, which we compute up to the quartic order in the conformal breaking scale. Our approach can be easily extended to discuss the 
anomaly contributions to the dilaton effective action to an arbitrarily high order. They allow to make a distinction between 
the Higgs and a dilaton at a phenomenological level.

\end{abstract}
\end{center}
 \newpage
\section{Introduction} 
The possibility that the Standard Model be characterized at high energy by a nearly conformal dynamics has motivated several investigations spanning considerable time \cite{Buchmuller:1987uc, Buchmuller:1988cj, Goldberger:2007zk, Shaposhnikov:2008xi, Meissner:2006zh}. If not for a quadratic term present in the Higgs potential, the model would in fact enjoy a dilatation symmetry which 
is broken by the vev of the Higgs field in the process of spontaneous symmetry breaking. 

A dilaton couples to the trace of the energy momentum tensor (EMT) ${T^\mu}_\mu$, and the coupling is affected by a trace anomaly. The trace anomaly equation plays a key role in characterizing the dynamics of the dilaton interactions, with a breaking of the dilatation symmetry which is enforced by two different contributions. 

They can be easily identified from the structure of the corresponding Ward identity satisfied by the 
dilaton $(\rho)VV$ vertex \cite{Coriano:2012nm}, with $V$ denoting a neutral (or charged) vector current, but also of higher vertices, such as the cubic and quartic 
dilaton interactions, which are part of the dilaton effective action. One specific contribution is the coupling of the dilaton to the anomaly, the second one 
being related to explicit mass terms generated at the electroweak scale.  In fact, the basic trace anomaly equation which takes the role of the generating functional of all the Ward identities satisfied by the dilaton vertices is given by 
\beqa 
\label{TraceAnomaly1}
g_{\mu\nu}(z)\left\langle T^{\mu\nu}(z) \right\rangle
&=& \mathcal A(z,g_{\alpha\beta})  +  \left\langle T^{\mu\nu}(z) \right\rangle\, ,
\eeqa
with $\left\langle \right\rangle$ denoting a functional average,  where the anomalous $(\mathcal A(z,g_{\alpha\beta}))$ and the explicit  $(\left\langle T^{\mu\nu}(z) \right\rangle)$ contributions are clearly separated. $g_{\alpha\beta}$ denotes the generic background metric which characterizes the anomalous contribution and which will be discussed in more detail below, when we will take the flat background limit. Gravity, in this case, plays simply an auxiliary role, since one takes the flat limit in all the hierarchical 
Ward identities which are obtained from (\ref{TraceAnomaly1}) after the functional differentiations. 

The goal of this work is to stress on some specific features of the dilaton effective action which follow up from (\ref{TraceAnomaly1}) and which are related to the anomalous structure of the anomalous contributions. In particular, in 
a nearly conformal phase of the Standard Model, which can be approximated by 
an exact $SU(3)\times SU(2)\times U(1)_Y$ gauge theory, cubic and quartic contributions to the dilaton dynamics are essentially fixed by the anomaly and can be extracted, with some effort, from a diagrammatic analysis of  (\ref{TraceAnomaly1}) expanded up to the fourth order in the metric.  This is the approach that we will be following in our case and on which we are going to elaborate. In particular we will present the expressions of such contributions. These interactions set a key distinction between a Higgs and a dilaton at all orders, being the Higgs not affected by the scale anomaly,  and can provide the basis for a direct phenomenological analysis of possible dilaton interactions at the LHC.

\section{Anomalous interactions from the Ward identities}
To illustrate the role of the anomaly in a more direct way and its possible significance in setting a distinction between the 
Higgs and the dilaton, we recall that the interaction Lagrangian of the dilaton $\rho$ with the Standard Model fields is given by
\beqa
\label{Lint}
\mathcal L_{int} = - \frac{1}{\Lambda_\rho} \rho \, {T^\mu}_\mu \,,
\eeqa
where $\Lambda_\rho$ is the conformal breaking scale which remains a free parameter of the effective action.
We introduce the ordinary definition of the energy-momentum tensor (EMT) of the Standard Model
\beq \label{EMT}
T^{\mu\nu}(z) = -\frac{2}{\sqrt{g_z}}\frac{\delta\,\mathcal{S}}{\delta g_{\mu\nu}(z)}  \, ,
\eeq
in terms of the quantum action $\mathcal{ S}$, so that its quantum average is
\beq \label{VEVEMT}
\left\langle T^{\mu\nu}(z) \right\rangle = \frac{2}{\sqrt{g_z}}\frac{\delta\, \mathcal W}{\delta\, g_{\mu\nu}(z)}\,,
\eeq
(with $\textrm{det}\, g_{\mu\nu}(z)\equiv g_z$) where $\mathcal W$ is the Euclidean generating functional of the theory.
$\mathcal W$ depends, in general, from the background metric $g_{\mu\nu}(x)$ defined as  
\beq\label{Generating}
\mathcal W = \frac{1}{\mathcal{N}} \, \int \, \mathcal D\Phi \, e^{-\, \mathcal{S}} \,. 
\eeq
\begin{figure}[t]
\label{rgg1}
\centering
\includegraphics[scale=0.8]{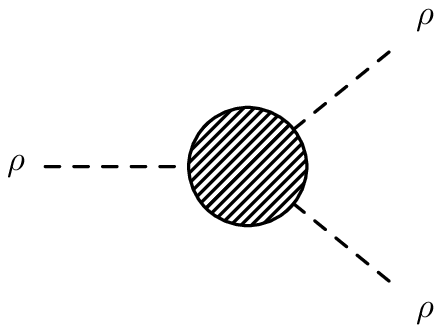}
\hspace{1cm}\includegraphics[scale=0.8]{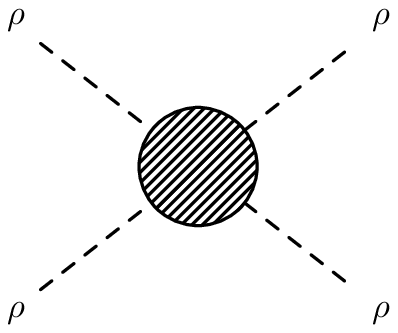}
\caption{Cubic and quartic dilaton interactions. In the nearly conformal limit the computation of the interactions involves virtual 
scalars, spin 1 and fermion exchanges. }
\end{figure}


The identification of the anomaly contributions to the dilaton interaction, on general grounds, requires an analysis of the anomalous Ward identities satisfied by the respective correlators. In this work we will concentrate on the extraction of the anomalous contribution to the quartic dilaton interactions, using as a fundamental scenario the Standard Model in the unbroken phase (i.e. with $v\to 0$). 

In this approximation  explicit trace insertions vanish for on-shell 
massless final states (i.e. for gauge fields), and we neglect all the contributions related both to the Higgs and to virtual corrections with a massive dilaton in the loops. In the same limit all the mixing contributions related to a possible term of improvement are not present \cite{Coriano:2012nm} and the computation of the anomalous terms amounts to the extraction of some finite parts.

Explicit (i.e. non anomalous) corrections, also present in the fundamental Ward identity, are calculable, but they are model dependent. In fact, they require the introduction of some extra potential for the dilaton/Higgs 
system. Its form has necessarily to rely on extra assumptions, such as the specific choice of breaking of the conformal symmetry. They are obtained by inserting the trace ($\bold{T}\equiv {T^\mu}_\mu$) operator on generic correlators involving all the fields of the Standard Model (plus other dilaton lines). In the conformal limit such contributions, which have been discussed in \cite{Coriano:2012nm} in the case of a single dilaton, drop out and the computations simplify considerably.

In this approximation the breaking of the dilatation symmetry does not contain any explicit scale-dependent term, and it is only due  to 
the anomaly, which is induced by renormalization. We call this approximation "nearly conformal".

As we have pointed out before \cite{Coriano:2012nm}, the breaking of the dilatation symmetry shows up, at a perturbative level, 
with the appearance of a massless pole in the $J_D VV$ correlator in the neutral and charged current sectors of the theory, with a residue which is 
proportional to a specific beta-function of the theory, related to the final state. This takes the role of a Nambu-Goldstone mode of the broken dilatation symmetry and it has been shown to affect each gauge invariant sector of the dilaton-to-two gauge bosons matrix elements.

To extract the anomalous contributions of the higher order interactions shown in Fig. 1 we start from the explicit expression of the anomaly, which  in $d=4$ is given by \cite{Duff:1977ay}, \cite{Duff:1993wm}

\beqa \label{TraceAnomaly}
 \mathcal A(z,g_{\alpha\beta})
&=&\sum_{I=f,s,V}n_I \bigg[\beta_a(I)\, F(z) + \beta_b(I)\, G(z) + \beta_c(I)\,\Box R(z) + \beta_d(I)\, R^2(z) \bigg].
\eeqa

$\mathcal{A}(z,g_{\alpha\beta})$ contains  the invariants built out of the Riemann tensor,
${R^{\alpha}}_{\beta\gamma\delta}$, as well as the Ricci tensor $R_{\alpha\beta}$ and the scalar curvature $R$.
G and $F$ in Eq. (\ref{TraceAnomaly}) are the Euler density and the square of the Weyl tensor respectively.\\ 
They are given by 
\beq\label{Geometry1}
F =
R^{\alpha\beta\gamma\delta}R_{\alpha\beta\gamma\delta} - 2\, R^{\alpha\beta}R_{\alpha\beta} + \frac{1}{3}R^2
\eeq
and
\beq\label{Euler}
G =
R^{\alpha\beta\gamma\delta}R_{\alpha\beta\gamma\delta} - 4\,R^{\alpha\beta}R_{\alpha\beta} + R^2\, ,
\eeq
with coefficients $\beta_a, \, \beta_b, \, \beta_c$ and $\beta_d$ which depend on the field content of the Lagrangian
(fermion, scalar, vector) and we have a multiplicity factor $n_I$ for each particle species. 

The coefficient of $R^2$ must vanish identically $(\beta_d \equiv 0)$ since a non-zero $R^2$ in this basis cannot be obtained from any effective action. In addition, the value of $\beta_c$ is regularization dependent, corresponding to 
the fact that it can be changed by the addition of an arbitrary local $R^2$ term in the effective action. The values given
in Table (\ref{AnomalyCoeff}) are those obtained in dimensional regularization, in which the relation $\beta_c = -{2}/{3}\,\beta_a$
is found to hold. Thus only $\beta_a$ and $\beta_b$  correspond to true anomalies in the trace of the EMT.
In Table \ref{AnomalyCoeff} we list the values of the coefficients for three theories of spin $0, \frac{1}{2}, 1$, that we are going to consider in the paper.
\begin{table}
$$
\begin{array}{|c|c|c|c|}\hline
I & \beta_a(I)\times 2880\,\pi^2 & \beta_b(I)\times 2880\,\pi^2 & \beta_c(I)\times 2880\,\pi^2
\\ \hline\hline
S & \frac{3}{2} & -\frac{1}{2} & -1
\\ \hline
F & 9 & -\frac{11}{2} & -6
\\ \hline
V & 18 & -31 & -12
\\
\hline
\end{array}
$$
\caption{Anomaly coefficients for a conformally coupled scalar, a Dirac Fermion and a vector boson}
\label{AnomalyCoeff}
\end{table}
In terms of the generating functional the fundamental trace anomaly equation can be rewritten in the form 
\beq
\label{TraceAnomalySymm}
2 \, g_{\mu\nu}(z)\frac{\delta\, \mathcal{W}}{\delta \, g_{\mu\nu}(z)}=
\sqrt{g_z} \, \mathcal A(z,g_{\alpha\beta}) \, 
\eeq
and plays the role of a generating functional for the anomalous Ward identities of any underlying 
Lagrangian field theory being, therefore, model independent. From (\ref{TraceAnomalySymm}) we can extract several identities satisfied by the anomaly term, for correlators involving $n$ 
insertions of energy momentum tensors, by performing $n$ functional derivatives with respect to the metric of both sides of 
(\ref{TraceAnomalySymm}) and taking the trace of the result at the end.

\section{EMT's and Correlators}
In perturbation theory, imposing the conservation Ward identity for the EMT and the Ward identities for the vector currents - whenever these are present - is sufficient to obtain
the corresponding anomalous term from the complete diagrammatic expansion. In particular, in dimensional regularization, the anomaly comes for free at the end of the computations, but this is a demanding job. 

In the case of the $TVV$,
for instance, it is a common practice to perform a direct computation, since only one term ($\sim F^{a\,\mu\nu}(z)\,F^a_{\mu\nu} (z)$)
can appear in the anomaly. We have omitted it in (\ref{TraceAnomaly}), since our analysis is focused on the anomaly-induced radiative corrections to correlators involving only dilaton self-interactions. A general discussion of contributions containing neutral currents (the $TVV$ vertex) has been given in \cite{Coriano:2011zk, Coriano:2012nm} in the Standard Model and in \cite{Giannotti:2008cv, Armillis:2009pq} for QED. However, things 
are far more involved for vertices containing multiple insertions of EMT's, such as the $TTT$ and $TTTT$, 
and it is convenient to infer the structure of the anomalous corrections without having to perform a complete 
diagrammatic analysis. In any case, a successful test of the anomalous Ward identities is crucial in order to secure the correctness of the result of the computation.

As mentioned above, in the nearly conformal limit of the Standard Model, we will need to consider a scalar, an abelian vector and fermion theory coupled to a background gravitational field. In fact the nonabelian character of the theory is not essential in the study of the higher order terms
to the dilaton effective action. In this case we can reconstruct the entire contribution to the anomaly from the abelian case by correcting the abelian result just by one extra multiplicity factor.

We will be using Euclidean conventions for the generating functional. The energy-momentum tensors for the theories that we consider are
\beqa
T^{\mu\nu}_{\phi}
&=&
\nabla^\mu \phi \, \nabla^\nu\phi - \frac{1}{2} \, g^{\mu\nu}\,g^{\alpha\beta}\,\nabla_\alpha \phi \, \nabla_\beta \phi
+ \chi \bigg[g^{\mu\nu} \Box - \nabla^\mu\,\nabla^\nu + \frac{1}{2}\,g^{\mu\nu}\,R - R^{\mu\nu} \bigg]\, \phi^2 
\label{ScalarEMT}\\
T^{\mu\nu}_{\psi}
&=&
\frac{1}{4} \,
\bigg[ g^{\mu\lambda}\,{V_\alpha}^\nu + g^{\nu\lambda}\,{V_\alpha}^\mu - 2\,g^{\mu\nu}\,{V_\alpha}^\lambda \bigg]
\bigg[\bar{\psi} \, \gamma^a \, \left(\mathcal{D}_\lambda \,\psi\right) -
\left(\mathcal{D}_\lambda \, \bar{\psi}\right) \, \gamma^a \, \psi \bigg] \, ,
\label{FermionEMT} \\
T^{\mu\nu}_A
&=&
F^{\mu\a}{F^\nu}_{\a}  - \frac{1}{4}g^{\mu\nu}F^{\a\b}F_{\a\b} \, ,
\label{MaxwellEMT}
\eeqa
where ${V_\alpha}^\nu$ is the \emph{vierbein} needed to embed the fermion in the gravitational background
and the corresponding covariant derivative is $ \mathcal{D}_\mu = \pd_\mu + \Gamma_\mu =
\pd_\mu + \frac{1}{2} \, \Sigma^{\alpha\beta} \, {V_\alpha}^\sigma \, \nabla_\mu\,V_{\beta\sigma} \, $
where the $\Sigma^{\alpha\beta}$ are the generators of the Lorentz group  in the case of a spin $1/2$-field. \\


It is convenient to define the correlation functions with $n$ external insertions of EMT's, which can be effectively thought as gravitons, as functional derivatives of order $n$ of $\mathcal W$, 
evaluated in the flat limit%
\beqa \label{NPF}
\left\langle T^{\mu_1\nu_1}(x_1)...T^{\mu_n\nu_n}(x_n) \right\rangle 
&=&
\bigg[\frac{2}{\sqrt{g_{x_1}}}...\frac{2}{\sqrt{g_{x_n}}} \,
\frac{\delta^n \mathcal{W}}{\delta g_{\mu_1\nu_1}(x_1)...\delta g_{\mu_n\nu_n}(x_n)}\bigg]
\bigg|_{g_{\mu\nu} = \delta_{\mu\nu}} \nonumber \\ 
&=&  
2^n\, \frac{\delta^n \mathcal{W}}{\delta g_{\mu_1\nu_1}(x_1)...\delta g_{\mu_n\nu_n}(x_n)}\bigg|_{g_{\mu\nu} = 
\delta_{\mu\nu}} \, .
\eeqa
We also define the notation
\beqa
\left[\mathcal F\right]^{\muu\nuu\mud\nud\dots\mu_{n}\nu_{n}}(\xu,\xd,\dots,x_n) \equiv
\frac{\delta^n\, \mathcal F}{\delta g_{\muu\nuu}(\xu)\,\delta g_{\mud\nud}(\xd)\dots\delta g_{\mu_n\nu_n}(x_{n})} 
\bigg|_{g_{\mu\nu}=\delta_{\mu\nu}}
\, , \label{Flat}
\eeqa
for any functional (or function) $\mathcal F$ which depends on the background field $g_{\mu\nu}(x)$. 
Denoting with \beq
\left\langle \mathcal{O} \right\rangle = \int\, \mathcal{D}\Phi\, \mathcal{O}\, e^{-S} \, 
\eeq
the vacuum expectation values of each operator, with $\mathcal{S}$ the generic action, we obtain 
\beqa
\left\langle T^{\mu\nu}(x)\right\rangle
&=&
- 2 \, \left\langle \left[\mathcal S\right]^{\mu\nu}(x) \right\rangle \label{1PF}\\
\left\langle T^{\muu\nuu}(x_1)T^{\mud\nud}(x_2)\right\rangle
&=&
4 \, \bigg[ 
\left\langle \left[\mathcal S\right]^{\muu\nuu}(\xu)\, \left[\mathcal S\right]^{\mud\nud}(\xd) \right\rangle 
- \, \left\langle \left[\mathcal S\right]^{\muu\nuu\mud\nud}(\xu,\xd)\right\rangle \bigg] \, , \label{2PF}\\
\left\langle T^{\muu\nuu}(x_1)T^{\mud\nud}(x_2)T^{\mut\nut}(x_3)\right\rangle
&=&
8 \, \bigg[ - \left\langle \left[\mathcal S \right]^{\muu\nuu}(\xu)\left[\mathcal S \right]^{\mud\nud}(\xd)
\left[\mathcal S \right]^{\mut\nut}(\xt) \right\rangle 
\nn \\
&& \hspace{-50mm}
+ \,\bigg( \left\langle \left[\mathcal S\right]^{\muu\nuu\mud\nud}(\xu,\xd)\,\left[\mathcal S\right]^{\mut\nut}(\xt) \right\rangle
+ 2\, \text{perm.} \bigg)
- \,\left\langle  \left[\mathcal S\right]^{\muu\nuu\mud\nud\mut\nut}(\xu,\xd,\xt)\right\rangle \bigg] \label{3PF} \, ,
\eeqa
\beqa
\left\langle T^{\muu\nuu}(x_1)T^{\mud\nud}(x_2)T^{\mut\nut}(x_3)T^{\muq\nuq}(x_4)\right\rangle
&=&
16 \, \bigg[
\left\langle \left[\mathcal S \right]^{\muu\nuu}(\xu)\left[\mathcal S \right]^{\mud\nud}(\xd)
\left[\mathcal S \right]^{\mut\nut}(\xt)\left[\mathcal S \right]^{\muq\nuq}(\xq) \right\rangle 
\nn \\
&& \hspace{-80 mm}
- \,\bigg( \left\langle \left[\mathcal S\right]^{\muu\nuu\mud\nud}(\xu,\xd)\left[\mathcal S\right]^{\mut\nut}(\xt)
\left[\mathcal S\right]^{\muq\nuq}(\xq) \right\rangle + 5 \, \text{perm.} \bigg) 
+ \bigg( \left\langle \left[\mathcal S\right]^{\muu\nuu\mud\nud}(\xu,\xd)\left[\mathcal 
S\right]^{\mut\nut\muq\nuq}(\xt,\xq)\right\rangle + 2 \, \text{perm.} \bigg)
\nn \\
&& \hspace{-80mm}
+ \,\bigg( \left\langle \left[\mathcal S\right]^{\muu\nuu\mud\nud\mut\nut}(\xu,\xd,\xt)\left[\mathcal 
S\right]^{\muq\nuq}(\xq)\right\rangle + 3 \, \text{perm.} \bigg)
- \, \left\langle \left[\mathcal S\right]^{\muu\nuu\mud\nud\mut\nut\muq\nuq}(\xu,\xd,\xt,\xq) \right\rangle \bigg] \, .
\label{4PF}
\eeqa
Notice that in dimensional regularization 
\beq
\left\langle \left[\mathcal S\right]^{\mu\nu}(x) \right\rangle= \left\langle \left[\mathcal 
S\right]^{\muu\nuu\mud\nud}(\xu,\xd)\right\rangle
=\left\langle  \left[\mathcal S\right]^{\muu\nuu\mud\nud\mut\nut}(\xu,\xd,\xt)\right\rangle
= \left\langle \left[\mathcal S\right]^{\muu\nuu\mud\nud\mut\nut\muq\nuq}(\xu,\xd,\xt,\xq) \right\rangle=0
 \eeq
being proportional to massless tadpoles. 
In particular, this implies that, to perform a perturbative computation of a correlator of order $n$,
one would be needing interaction vertices with at most $n-1$ gravitons.
Concerning the diagrammatic structure of each contribution, the correlator
\beq
\left\langle \left[\mathcal S \right]^{\muu\nuu}(\xu)\left[\mathcal S \right]^{\mud\nud}(\xd)
      \left[\mathcal S \right]^{\mut\nut}(\xt)\left[\mathcal S \right]^{\muq\nuq}(\xq) \right\rangle 
\eeq 
has a box topology;
\beq     
\left\langle \left[\mathcal S \right]^{\muu\nuu}(\xu)\left[\mathcal S \right]^{\mud\nud}(\xd)
\left[\mathcal S \right]^{\mut\nut}(\xt) \right\rangle 
\eeq
which is the first contribution to the graviton 3-point function, and
\beq 
\left\langle \left[\mathcal S\right]^{\muu\nuu\mud\nud}(\xu,\xd) 
\left[\mathcal S\right]^{\mut\nut}(\xt)\left[\mathcal S\right]^{\muq\nuq}(\xq) \right\rangle \, ,
\eeq 
which corresponds to a contact term in $\left\langle TTTT \right\rangle$, are represented by triangles. \\
The remaining contributions, 
 \beq
 \left\langle \left[\mathcal S\right]^{\muu\nuu}(\xu)\, \left[\mathcal S\right]^{\mud\nud}(\xd) 
      \right\rangle \, ,
\eeq 
the contact terms $\left\langle TTT \right\rangle$, which are 
\beq
\left\langle \left[\mathcal S\right]^{\muu\nuu\mud\nud}(\xu,\xd)\,\left[\mathcal S\right]^{\mut\nut}(\xt) \right\rangle
\eeq 
and the two remaining types of diagrams which enter into $\left\langle TTTT \right\rangle$,
\beq     
\left\langle \left[\mathcal S\right]^{\muu\nuu\mud\nud}(\xu,\xd)\left[\mathcal S\right]^{\mut\nut\muq\nuq}(\xt,\xq)\right\rangle
\eeq
and
\beq
     \left\langle \left[\mathcal S\right]^{\muu\nuu\mud\nud\mut\nut}(\xu,\xd,\xt)\left[\mathcal 
      S\right]^{\muq\nuq}(\xq)\right\rangle \, ,
\eeq
have the topologies of 2-point functions.
Our conventions for the choice of the external momenta are defined via the Fourier transforms
\beqa
&&
\int \, d^4\xu\,d^4\xd\,d^4\xt\,d^4\xq\, 
\left\langle T^{\muu\nuu}(\xu)T^{\mud\nud}(\xd)T^{\mut\nut}(\xt)T^{\muq\nuq}(\xq)\right\rangle \,
e^{-i(\ku\cdot \xu + \kd\cdot \xd + \kt\cdot \xt + \kq \cdot \xq)} = 
\nn \\
&& \hspace{40mm}
(2\pi)^4\,\delta^{(4)}(\ku + \kd + \kt + \kq)\, \left\langle T^{\muu\nuu}T^{\mud\nud}T^{\mut\nut}T^{\muq\nuq}\right\rangle(\kd,\kt,\kq) 
\, , 
\label{4PFMom}
\eeqa
and similar for the 3- and 2-point functions. We will be taking all momenta to be incoming.

\section{Ward identities}


We start from the analysis of the general covariance Ward identites. These  are obtained from the functional relation 
\beq\label{masterWI0}
\nabla_{\nuu} \left\langle T^{\muu\nuu}(x_1) \right\rangle
= \nabla_{\nuu} \bigg(\frac{2}{\sqrt{g_{x_1}}}\frac{\delta\mathcal W}{\delta g_{\muu\nuu}(x_1)}\bigg)
= 0 \, 
\eeq
which after an expansion becomes
\beqa \label{InterWard}
&&
\frac{2}{\sqrt{g_{x_1}}}\bigg(\pd_{\nuu}\frac{\delta\mathcal W}{\delta g_{\muu\nuu}(x_1)}
- \Gamma^\lambda_{\lambda\nuu}(x_1)\frac{\delta\mathcal W}{\delta g_{\muu\nuu}(x_1)}
+ \Gamma^{\muu}_{\kappa\nuu}(x_1)\frac{\delta\mathcal W}{\delta g_{\kappa\nuu}(x_1)}
+ \Gamma^{\nuu}_{\kappa\nuu}(x_1)\frac{\delta\mathcal W}{\delta g_{\muu\kappa}(x_1)}\bigg) = 0. \nn\\
\eeqa
Cancelling the secod and fourth terms in parentheses, (\ref{InterWard}) takes the simpler form
\beqa \label{masterWI}
&& 
2 \, \bigg(\pd_{\nuu}\frac{\delta\mathcal W}{\delta g_{\muu\nuu}(x_1)}
+ \Gamma^{\muu}_{\kappa\nuu}(x_1)\frac{\delta\mathcal W}{\delta g_{\kappa\nuu}(x_1)}\bigg)
= 0\, .
\eeqa
The Ward identities we are interested in are obtained by functional differentiation of (\ref{masterWI}) and give
\beqa
&&
4\, \bigg[ \pd_{\nuu} \frac{\delta^2\mathcal W}{\delta g_{\mud\nud}(x_2)\delta g_{\muu\nuu}(x_1)}
+ \frac{\delta \Gamma^{\muu}_{\kappa\nuu}(x_1)}{\delta g_{\mud\nud}(x_2)}\frac{\delta\mathcal W}{\delta g_{\kappa\nuu}(x_1)}
+ \Gamma^{\muu}_{\kappa\nuu}(x_1) \frac{\delta^2 \mathcal W}{\delta g_{\mud\nud}(x_2)\delta g_{\kappa\nuu}(x_1)}\bigg]
= 0, \, \label{WI2PFCoordinate} 
\eeqa
for single
\beqa
&&
8\, \bigg[\pd_{\nuu}\frac{\delta^3\mathcal W}{\delta g_{\mut\nut}(x_3)\delta g_{\mud\nud}(x_2)\delta g_{\muu\nuu}(x_1)}
+ \frac{\delta \Gamma^{\muu}_{\kappa\nuu}(x_1)}{\delta g_{\mud\nud}(x_2)}
\frac{\delta^2 \mathcal W}{\delta g_{\mut\nut}(x_3)\delta g_{\kappa\nuu}(x_1)}
+ \frac{\delta \Gamma^{\muu}_{\kappa\nuu}(x_1)}{\delta g_{\mut\nut}(x_3)}
\frac{\delta^2 \mathcal W}{\delta g_{\mud\nud}(x_2)\delta g_{\kappa\nuu}(x_3)} \nn\\
&&
+ \frac{\delta^2\Gamma^{\muu}_{\kappa\nuu}(x_1)}{\delta g_{\mud\nud}(x_2)\delta g_{\mut\nut}(x_3)}
\frac{\delta\mathcal W}{\delta g_{\kappa\nuu}(x_1)}
+ \Gamma^{\muu}_{\kappa\nuu}(x_1)
\frac{\delta^3\mathcal W}{\delta g_{\mud\nud}(x_2)\delta g_{\mut\nut}(x_2)\delta g_{\kappa\nuu}(x_1)}\bigg] = 0 \, ,
\label{WI3PFCoordinate} 
\eeqa
double, and 
\beqa
&&
16\, \bigg[\pd_{\nuu}\frac{\delta^4\mathcal W}
{\delta g_{\muq\nuq}(x_4)\delta g_{\mut\nut}(x_3)\delta g_{\mud\nud}(x_2)\delta g_{\muu\nuu}(x_1)}
+ \bigg(\frac{\delta \Gamma^{\muu}_{\kappa\nuu}(x_1)}{\delta g_{\mud\nud}(x_2)}
\frac{\delta^3 \mathcal W}{\delta g_{\muq\nuq}(x_4) \delta g_{\mut\nut}(x_3)\delta g_{\kappa\nuu}(x_1)}
\nn \\
&&
+ \big( 2 \leftrightarrow 4, 2 \leftrightarrow 3\big) \bigg)
+ \bigg( \frac{\delta^2 \Gamma^{\muu}_{\kappa\nuu}(x_1)}{\delta g_{\mut\nut}(x_3) \delta g_{\mud\nud}(x_2)}
\frac{\delta^2 \mathcal W}{\delta g_{\muq\nuq}(x_4)\delta g_{\kappa\nuu}(x_1)} + 
\big( 2 \leftrightarrow 4, 3 \leftrightarrow 4 \big) \bigg)
\nn \\
&&
+ \frac{\delta^3 \Gamma^{\muu}_{\kappa\nuu}(x_1)}{\delta g_{\muq\nuq}(x_4) \delta g_{\mut\nut}(x_3) \delta g_{\mud\nud}(x_2)}
\frac{\delta \mathcal W}{\delta g_{\kappa\nuu}(x_1)}
+ \Gamma^{\muu}_{\kappa\nuu}(x_1) \frac{\delta^4 \mathcal W}
{\delta g_{\muq\nuq}(x_4)\delta g_{\mut\nut}(x_3)\delta g_{\mud\nud}(x_2)\delta g_{\kappa\nuu}(x_1)}=0
\label{WI4PFCoordinate} 
\eeqa
triple differentiations of the master equation (\ref{masterWI0}).
In the Ward identity satisfied by the 4-point function we have left implicit the contributions obtained by permuting the action of 
the functional derivatives. 

To move to the flat spacetime limit of (\ref{WI2PFCoordinate}) and (\ref{WI3PFCoordinate}), 
we use the notations in (\ref{Flat}) and set to zero the contributions from the massless tadpoles, obtaining  
\beqa
\pd_{\nuu} \left\langle T^{\muu\nuu}(x_1) T^{\mud\nud}(x_2) \right\rangle &=& 0 \, , \label{WI2PFCoordinateFlat} \\
\pd_{\nuu} \left\langle T^{\muu\nuu}(x_1)T^{\mud\nud}(x_2)T^{\mut\nut}(x_3) \right\rangle 
&=& 
- 2\, \left[\Gamma^{\muu}_{\kappa\nuu}(x_1)\right]^{\mud\nud}(x_2) \left\langle T^{\kappa\nuu}(x_1)T^{\mut\nut}(x_3)\right\rangle
\nn \\
&&
- 2\, \left[\Gamma^{\muu}_{\kappa\nuu}(x_1)\right]^{\mut\nut}(x_3) \left\langle T^{\kappa\nuu}(x_1)T^{\mut\nut}(x_3)\right\rangle 
\, , \label{WI3PFCoordinateFlat} \\
\pd_{\nuu} \left\langle T^{\kappa\nuu}(x_1)T^{\mud\nud}(x_2)T^{\mut\nut}(x_3)T^{\muq\nuq}(x_4) \right\rangle 
&=&
- 2\, \bigg(\left[\Gamma^{\muu}_{\kappa\nuu}(x_1)\right]^{\mud\nud}(x_2)
\left\langle T^{\kappa\nuu}(x_1) T^{\mut\nut}(x_3)T^{\muq\nuq}(x_4) \right\rangle 
\nn \\
&& \hspace{-40mm}
+ \big( 2 \leftrightarrow 3, 2 \leftrightarrow 4 \big) \bigg)
- 4\, \bigg( \left[\Gamma^{\muu}_{\kappa\nuu}(x_1)\right]^{\mud\nud\mut\nut}(x_2,x_3)
\left\langle T^{\kappa\nuu}(x_1)T^{\muq\nuq}(x_4) \right\rangle 
+ \big( 2 \leftrightarrow 4, 3 \leftrightarrow 4 \big) \bigg) \, , \label{WI4PFCoordinateFlat} 
\nn \\
\eeqa
which after some manipulations give the transversality constraint for the 2-point functions and
\beqa
k_{1\,\nuu} \left\langle T^{\muu\nuu}T^{\mut\nut}T^{\mud\nud} \right\rangle(\kd,\kt) 
&=&
- \kt^{\muu} \left\langle T^{\mut\nut}T^{\mud\nud}\right\rangle(\kd) 
- \kd^{\muu} \left\langle T^{\mud\nud}T^{\mut\nut}\right\rangle(\kt) \nn \\
&+&
k_{3\, \nuu} \bigg[\delta^{\muu\nut} \left\langle T^{\nuu\mut}T^{\mud\nud} \right\rangle(\kd)
+ \delta^{\muu\mut}\left\langle T^{\nuu\nut}T^{\mud\nud}  \right\rangle(\kd)\bigg] 
\nn \\
&+& 
k_{2\, \nuu} \bigg[\delta^{\muu\nud}\left\langle T^{\nuu\mud}T^{\mut\nut}  \right\rangle(\kt)
+ \delta^{\muu\mud}  \left\langle T^{\nuu\nud}T^{\mut\nut}\right\rangle(\kt)\bigg]	\, .
\nn \\
\label{WI3PF}
\eeqa
Similarly, in the case of the 4-point function $TTTT$, using (\ref{4PFMom}) we obtain
\beqa
&&
k_{1\,\nuu} \, \left\langle T^{\muu\nuu}T^{\mut\nut}T^{\mud\nud}T^{\muq\nuq} \right\rangle(\kd,\kt,\kq) =
\bigg[ - \kd^{\muu} \, \left\langle T^{\mud\nud}T^{\mut\nut}T^{\muq\nuq}\right\rangle(\kt,\kq)
\nn \\
&+&
 k_{2\,\nuu}\, \bigg( \delta^{\muu\nud}\, \left\langle T^{\nuu\mud}T^{\mut\nut}T^{\muq\nuq}\right\rangle(\kt,\kq)
+ \delta^{\muu\mud}\, \left\langle T^{\nuu\nud}T^{\mut\nut}T^{\muq\nuq}\right\rangle(\kt,\kq) \bigg)
+ \big( 2 \leftrightarrow 3, 2 \leftrightarrow 4\big) \bigg]
\nn \\
&+&
 \bigg[ 2\, k_{2\,\nuu}\, \bigg( 
\left[g^{\muu\mud}\right]^{\mut\nut}\, \left\langle T^{\nuu\nud} T^{\muq\nuq}\right\rangle(\kq) + 
\left[g^{\muu\nud}\right]^{\mut\nut}\, \left\langle T^{\nuu\mud}T^{\muq\nuq} \right\rangle(\kq) \bigg)
\nn \\
&+& \hspace{2mm}
 2\, k_{3\,\nuu}\, \bigg(
\left[g^{\muu\mut}\right]^{\mud\nud}\, \left\langle T^{\nuu\nut} T^{\muq\nuq} \right\rangle(\kq) + 
\left[g^{\muu\nut}\right]^{\mud\nud}\, \left\langle T^{\nuu\mut} T^{\muq\nuq} \right\rangle(\kq) \bigg)
\nn \\
&+&
 \bigg( k_2^{\nut} \delta^{\muu\mut} + k_2^{\mut} \delta^{\muu\nut}\bigg)\, \left\langle T^{\mud\nud}T^{\mut\nut}\right\rangle(\kq)
+ \bigg( k_3^{\nud} \delta^{\muu\mud} + k_3^{\mud} \delta^{\muu\nud}\bigg)\, \left\langle T^{\mut\nut}T^{\mud\nud}\right\rangle(\kq)
+ \big( 2 \leftrightarrow 4, 3 \leftrightarrow 4 \big) \bigg].
\nn \\
\label{WI4PF}
\eeqa
Similar identities are obtained for the momenta of the other external gravitons. 

\section{Counterterms}

Coming to a discussion of the counterterms to the 4-dilaton amplitude, these are obtained from the one-loop Lagrangian 
which accounts for the gravitational counterterms to pure graviton amplitudes
\beq\label{CounterAction}
S_{counter} = - \frac{1}{\epsilon}\sum_{I=f,s,V}n_I \int d^d x \sqrt{g} \bigg( \beta_a(I) F + \beta_b(I) G\bigg),
\eeq
with $\epsilon= 4 - d$, containing the squared Weyl tensor $F$ and the Euler density, defined above.
In the case of the 4-graviton vertex the counterterm action (\ref{CounterAction}) generates the vertices 
\beq
-\frac{1}{\epsilon}\, \bigg(
\beta_a\, D_F^{\muu\nuu\mud\nud\mut\nut\muq\nuq}(x_1,x_2,x_3,x_4) + 
\beta_b\, D_G^{\muu\nuu\mud\nud\mut\nut\muq\nuq}(x_1,x_2,x_3,x_4)\bigg)\, ,
\eeq
where
\beqa
D_F^{\muu\nuu\mud\nud\mut\nut\muq\nuq}(x_1,x_2,x_3,x_4)
&=&
2^4 \, \frac{\delta^4}{\delta g_{\muu\nuu}(x_1)\delta g_{\mud\nud}(x_2)\delta g_{\mut\nut}(x_3)\delta g_{\muq\nuq}(x_4)}
\int\,d^d w\,\sqrt{g}\, F\, ,
\label{DF}\\
D_G^{\muu\nuu\mud\nud\mut\nut\muq\nuq}(x_1,x_2,_3,x_4)
&=&
2^4 \, \frac{\delta^4}{\delta g_{\muu\nuu}(x_1)\delta g_{\mud\nud}(x_2)\delta g_{\mut\nut}(x_3)\delta g_{\muq\nuq}(x_4)}
\int\,d^d w\,\sqrt{g}\, G 
\label{DG}\,, 
\eeqa
and similarly for the 2- and 3-point correlators.
The functional derivation of such integrals, limited to 3-point functions, has been detailed in \cite{Coriano:2012wp}.
Using these expressions, the fully renormalized 2-, 3- and 4-point correlators in momentum space can be written down as
\beqa
\left\langle T^{\muu\nuu}T^{\mud\nud} \right\rangle_{ren}(\kd) 
&=&
\left\langle T^{\muu\nuu}T^{\mud\nud} \right\rangle_{bare}(\kd) -
\frac{1}{{\epsilon}}\, \beta_a\, D_F^{\muu\nuu\mud\nud}(\kd) \, ,
\label{Ren2PF}
\nn \\
\left\langle T^{\muu\nuu}T^{\mud\nud}T^{\mut\nut} \right\rangle_{ren}(\kd,\kt)
&=&
\left\langle T^{\muu\nuu}T^{\mud\nud}T^{\mut\nut} \right\rangle_{bare}(\kd,\kt)
\nn \\
&-&
\frac{1}{{\epsilon}}\,\bigg(
\beta_a\, D_F^{\muu\nuu\mud\nud\mut\nut}(\kd,\kt) + 
\beta_b\, D_G^{\muu\nuu\mud\nud\mut\nut}(\kd,\kt) \bigg)\, ,
\label{Ren3PF}
\nn \\
\left\langle T^{\muu\nuu}T^{\mud\nud}T^{\mut\nut}T^{\muq\nuq} \right\rangle_{ren}(\kd,\kt,\kq)
&=&
\left\langle T^{\muu\nuu}T^{\mud\nud}T^{\mut\nut}T^{\muq\nuq} \right\rangle_{bare}(\kd,\kt,\kq)
\nn \\
&-&
\frac{1}{{\epsilon}}\,\bigg(
\beta_a\, D_F^{\muu\nuu\mud\nud\mut\nut\muq\nuq}(\kd,\kt,\kq) + 
\beta_b\, D_G^{\muu\nuu\mud\nud\mut\nut\muq\nuq}(\kd,\kt,\kq) \bigg)\,  .
\nn \\
\label{Ren4PF}
\eeqa
From these relations and from (\ref{WI3PF}), (\ref{WI4PF}) it is clear that counterterms must be related by the same general covariance
Ward identities which relate the bare correlators.  One can also separately check these identites for the F- and G- counterterms just 
by writing them down and equating the coefficients of $\beta_a$ and $\beta_b$.
We omit the explicit forms of the counterterms, which are necessary in order to test all these constraints, that we have checked for consistency.

\subsection{Anomalous trace identities}

Anomalous Ward identities for the Green functions at hand are obtained through the functional differentiation of (\ref{TraceAnomalySymm}), 
passing to the flat space limit and using the definition (\ref{NPF}). A complex computation gives
\beqa 
\delta_{\muu\nuu}\, \left\langle T^{\muu\nuu}T^{\mud\nud}\right\rangle(\kd)
&=&
2 \, \left[\sqrt{g}\, \mathcal{A}\right]^{\mud\nud}(\kd) = 2 \, \left[ \Box R\right]^{\mud\nud}(\kd)\, , \label{AWard2PF} \\
\delta_{\muu\nuu}\, \left\langle T^{\muu\nuu}T^{\mud\nud}T^{\mut\nut} \right\rangle(\kd,\kt)
&=&
4 \, \left[\sqrt{g}\, \mathcal{A}\right]^{\mud\nud\mut\nut}(\kd,\kt)
- 2 \, \left\langle T^{\mud\nud}T^{\mut\nut} \right\rangle(\kd) 
- 2 \, \left\langle T^{\mud\nud}T^{\mut\nut} \right\rangle(\kt)\nn\\
&& \hspace{-40mm}
=\, 4 \, \bigg[ \beta_a\,\bigg(\left[F\right]^{\mut\nut\mud\nud}(\kd,\kt)
- \frac{2}{3} \left[\sqrt{g}\, \Box R\right]^{\mut\nut\mud\nud}(\kd,\kt)\bigg)
+ \beta_b\, \left[G\right]^{\mut\nut\mud\nud}(\kd,\kt) \bigg]\nn\\
&-&
  2 \, \left\langle T^{\mud\nud}T^{\mut\nut} \right\rangle(\kd) 
- 2 \, \left\langle T^{\mud\nud}T^{\mut\nut} \right\rangle(\kt) \, , \label{AWard3PF} \\
\delta_{\muu\nuu}\, \left\langle T^{\muu\nuu}T^{\mud\nud}T^{\mut\nut}T^{\muq\nuq} \right\rangle(\kd,\kt,\kq)
&=&
8 \, \left[ \sqrt{g}\, \mathcal{A}\right]^{\mud\nud\mut\nut\muq\nuq}(\kd,\kt,\kq)
\nn \\
&& \hspace{-60mm}
-\,  2 \, \left\langle T^{\mud\nud}T^{\mut\nut}T^{\muq\nuq} \right\rangle(\kd,\kt)
- 2 \, \left\langle T^{\mud\nud}T^{\mut\nut}T^{\muq\nuq} \right\rangle(\kd,\kq)
- 2 \, \left\langle T^{\mud\nud}T^{\mut\nut}T^{\muq\nuq} \right\rangle(\kt,\kq) 
\nn\\
&& \hspace{-70mm}
=\, 8 \, \bigg[ \beta_a\, \bigg(\left[\sqrt{g}\,F\right]^{\mud\nud\mut\nut\muq\nuq}(\kd,\kt,\kq)
- \frac{2}{3} \left[\sqrt{g}\, \Box R\right]^{\mud\nud\mut\nut\muq\nuq}(\kd,\kt,\kq)\bigg)
+ \beta_b\, \left[\sqrt{g}\, G\right]^{\mud\nud\mut\nut\muq\nuq}(\kd,\kt,\kq) \bigg] 
\nn\\
&& \hspace{-70mm}
- 2 \, \left\langle T^{\mud\nud}T^{\mut\nut}T^{\muq\nuq} \right\rangle(\kd,\kt)
- 2 \, \left\langle T^{\mud\nud}T^{\mut\nut}T^{\muq\nuq} \right\rangle(\kd,\kq)
- 2 \, \left\langle T^{\mud\nud}T^{\mut\nut}T^{\muq\nuq} \right\rangle(\kt,\kq) \, . \label{AWard4PF}
\eeqa
The explicit expressions of the multiple functional derivatives of the various operators in square bracket $([\,\,])$ 
are very lengthy and we omit them.

At this stage we can extract from (\ref{AWard4PF}) four trace-identities
(one for each graviton) for the counterterms of the 4-point functions  (\ref{DF}) and (\ref{DG}), relating them to their corresponding 
2- and 3-point ones. All these counterterms have been indipendently tested through general covariance Ward identities. This approach allows to check also the structure of the anomaly contributions to the 4-point function, and in fact it is used to deduce the form of the quartic dilaton 
interactions. The identity is
\beqa
\delta^{d}_{\muu\nuu}\, D_{F}^{\muu\nuu\mud\nud\mut\nut\muq\nuq}(\kd,\kt,\kq) 
&=& 
- \frac{\epsilon}{2}\, \left(\left[\sqrt{g}\, F\right]^{\mud\nud\mut\nut\muq\nuq}(\kd,\kt,\kq) 
- \frac{2}{3}\,  \left[\sqrt{g}\,\Box R \right]^{\mud\nud\mut\nut\muq\nuq}(\kd,\kt,\kq)\right)
\nn \\
&-& 
2\, D_{F}^{\mud\nud\mut\nut\muq\nuq}(\kt,\kq) - 2\, D_{F}^{\mud\nud\mut\nut\muq\nuq}(\kd,\kq)
- 2\, D_{F}^{\mud\nud\mut\nut\muq\nuq}(\kd,\kt) \, , 
\nn \\
\delta^{d}_{\muu\nuu}\, D_{F}^{\muu\nuu\mud\nud\mut\nut\muq\nuq}(\kd,\kt,\kq) 
&=& 
- \frac{\epsilon}{2}\, \left[\sqrt{g}\, G \right]^{\mud\nud\mut\nut\muq\nuq}(\kd,\kt,\kq)
\nn\\
&-& 
2\, D_{G}^{\mud\nud\mut\nut\muq\nuq}(\kt,\kq) - 2\, D_{G}^{\mud\nud\mut\nut\muq\nuq}(\kd,\kq)
- 2\, D_{G}^{\mud\nud\mut\nut\muq\nuq}(\kd,\kt) \, , 
\nn \\
\eeqa
where the superscript $d$ is there to indicate that the trace has to be taken in $d = 4 - \epsilon$ dimensions.

\subsection{Three and four dilaton interactions from the trace anomaly}

From (\ref{AWard2PF})-(\ref{AWard4PF}) and from the knowledge of the trace anomalies therein, that we have explicitly computed,
one can get the form of the off-shell 3- and 4-dilaton ($\rho$) interactions,
which are found to be
%
\beqa \label{3DVertex}
\mathcal V_{\rho\rho\rho}^{\phi}
&=&
- \frac{1}{\Lambda_\rho^3}\, \frac{3\,\left( \kd^4+ \kt^4\right) + 6 \, \kd\cdot \kt \, \left( \kd^2 + \kt^2 \right) 
+ 4 \,(\kd\cdot \kt)^2 + 5 \, \kd^2\,\kt^2}{360\,\pi^2} 
\, , \nonumber \\
\mathcal V_{\rho\rho\rho}^{\psi}
&=&
- \frac{1}{\Lambda_\rho^3}\, \frac{18\,\left( \kd^4+ \kt^4\right) + 36 \, \kd\cdot \kt \, \left( \kd^2 + \kt^2 \right) 
+ 29 \,(\kd\cdot \kt)^2 + 25 \, \kd^2\,\kt^2}{360\, \pi^2} \, , 
\nonumber \\
\mathcal V_{\rho\rho\rho}^A
&=&
- \frac{1}{\Lambda_\rho^3}\, \frac{18\,\left( \kd^4+ \kt^4\right) + 36 \, \kd\cdot \kt \, \left( \kd^2 + \kt^2 \right) 
+ 49 \,(\kd\cdot \kt)^2 + 5 \, \kd^2\,\kt^2}{180\,\pi^2}
\eeqa
in agreement with the result given in \cite{Coriano:2012nm}
and the new quartic dilaton interactions 
\beqa \label{4DVertex}
\mathcal V_{\rho\rho\rho\rho}^{\phi}
&=&
- \frac{1}{\Lambda_\rho^4}\,\frac{1}{60\,\pi^2}\, \bigg[
3\, \left((\kd^2)^2 + (\kt^2)^2 + (\kq^2)^2 \right) + 
6\, \left( \kq^2\, \kq \cdot (\kd +\kt) +  \kt^2\, \kt \cdot(\kd + \kq) +  \kd^2\, \kd\cdot(\kt+\kq) \right)
\nn \\
&& \hspace{-10mm}
+ \, 4\, \left( ( \kd \cdot \kq)^2 + ( \kd \cdot \kt)^2 + ( \kt \cdot \kq)^2 \right) + 
6\, \left( \kd \cdot \kt\, \kd \cdot \kq + \kt \cdot \kd\, \kt \cdot \kq +  \kq \cdot \kd\, \kq \cdot \kt  \right)
\nn \\
&& \hspace{-10mm}
+\, 5\, \left( \kd^2\, \kt^2 + \kt^2\, \kq^2 +  \kd^2\, \kq^2 \right)
+ 5\, \left( \kd^2\, \kt \cdot \kq + \kt^2\, \kd \cdot \kq +  \kq^2\, \kd \cdot \kt \right) \bigg] \, , 
\nonumber \\
\mathcal V_{\rho\rho\rho\rho}^{\psi}
&=&
- \frac{1}{\Lambda_\rho^4}\, \frac{1}{120\,\pi^2}\,
\, \bigg[
36\, \left((\kd^2)^2 + (\kt^2)^2 + (\kq^2)^2 \right) + 
72\, \left( \kq^2\, \kq \cdot (\kd +\kt) +  \kt^2\, \kt \cdot(\kd + \kq) +  \kd^2\, \kd\cdot(\kt+\kq) \right)
\nn \\
&& \hspace{-10mm}
+ \, 58\, \left( ( \kd \cdot \kq)^2 + ( \kd \cdot \kt)^2 + ( \kt \cdot \kq)^2 \right) + 
82\, \left( \kd \cdot \kt\, \kd \cdot \kq + \kt \cdot \kd\, \kt \cdot \kq +  \kq \cdot \kd\, \kq \cdot \kt  \right)
\nn \\
&& \hspace{-10mm}
+\, 50\, \left( \kd^2\, \kt^2 + \kt^2\, \kq^2 +  \kd^2\, \kq^2 \right)
+ 55\, \left( \kd^2\, \kt \cdot \kq + \kt^2\, \kd \cdot \kq +  \kq^2\, \kd \cdot \kt \right) \bigg] \, , 
\nonumber \\
\mathcal V_{\rho\rho\rho\rho}^A
&=&
- \frac{1}{\Lambda_\rho^4}\, \frac{1}{60\,\pi^2}\,
\, \bigg[
36\, \left((\kd^2)^2 + (\kt^2)^2 + (\kq^2)^2 \right) + 
72\, \left( \kq^2\, \kq \cdot (\kd +\kt) +  \kt^2\, \kt \cdot(\kd + \kq) +  \kd^2\, \kd\cdot(\kt+\kq) \right)
\nn \\
&& \hspace{-10mm}
+ \,98 \, \left( ( \kd \cdot \kq)^2 + ( \kd \cdot \kt)^2 + ( \kt \cdot \kq)^2 \right) + 
122\, \left( \kd \cdot \kt\, \kd \cdot \kq + \kt \cdot \kd\, \kt \cdot \kq +  \kq \cdot \kd\, \kq \cdot \kt  \right)
\nn \\
&& \hspace{-10mm}
+\,10 \, \left( \kd^2\, \kt^2 + \kt^2\, \kq^2 +  \kd^2\, \kq^2 \right)
+ 35 \, \left( \kd^2\, \kt \cdot \kq + \kt^2\, \kd \cdot \kq +  \kq^2\, \kd \cdot \kt \right) \bigg] \, .
\eeqa
Both the cubic and the quartic terms, in the nearly conformal limit, can be easily modified to account for all the contributions generated in the case of Standard Model in the form
\bea
\mathcal V_{\rho\rho\rho / \rho \rho\rho\rho} = \sum_{i=A,\psi,\phi} N_i \, V^i_{\rho\rho\rho / \rho \rho\rho\rho}
\eea
where $N_A = 8+3+1 = 12$ is the number of gauge fields, $N_\psi = 3 \times 6 + 3 + 3/2 = 45/2$ is the number of Dirac fermions, where the factor 1/2 is due to the fermion chirality, and $N_\phi = 4$ counts the real scalars of the Higgs doublet. 
These corrections, as we have already remarked, are typical of the dilaton interactions and can be derived 
without any explicit diagrammatic computation. They provide the starting ground for an analysis - up to an arbitrarily high order - of the dilaton effective action, and characterize the terms which allows to differentiate between the Higgs and the dilaton at the radiative level.

\section{Conclusions} 
The analysis of dilaton interactions and of their role in the context of the electroweak symmetry breaking is particularly interesting at phenomenological level, especially since, at this time, there is  not yet a complete understanding of the properties of the 126 GeV scalar resonance in the diphoton channel, recently detected at the LHC by ATLAS and CMS. 
In fact, the Standard Model, in the limit in which we drop the Higgs mass, is conformally invariant at high energy, with a breaking of scale 
invariance, in this limit, which is related only to the trace anomaly, because of renormalization. 
The anomalous coupling of the dilaton is responsible for setting a remarkable difference between this state and the Higgs, a property which remains valid - with no distinction - even if the dilaton is assumed to be a fundamental or a composite scalar or a graviscalar.  We have shown that the anomalous corrections at any order to the dilaton effective action, in the conformal limit, can be extracted from a general (and model independent) analysis of the Ward identities, with no further input. We have illustrated the approach up to the quartic order.
\vspace{1cm}

\centerline{\bf Acknowledgements} 
We thank Roberta Armillis and Abdelhak Djouadi for discussions.


\end{document}